\documentclass[aps,prd,twocolumn,superscriptaddress,groupedaddress,nofootinbib]{revtex4}  
\usepackage{graphicx}  
\usepackage{dcolumn}   
\usepackage{bm}        
\usepackage{amssymb}   
\usepackage{hyperref}   

\begin{document}



\title{Di-photon decay of a light Higgs state in the BLSSM}
\author{Ahmed Ali Abdelalim}
\email{aabdelalim@zewailcity.edu.eg}
\affiliation{Center for Fundamental Physics, Zewail City of Science and Technology, \\
6 October City, Giza 12588, Egypt}
\author{Biswaranjan Das} 
\email{bdas@zewailcity.edu.eg}
\affiliation{Center for Fundamental Physics, Zewail City of Science and Technology, \\
6 October City, Giza 12588, Egypt}
\affiliation{East African Institute for Fundamental Research (ICTP-EAIFR), \\
University of Rwanda, Kigali, Rwanda}
\author{Shaaban Khalil} 
\email{skhalil@zewailcity.edu.eg}
\affiliation{Center for Fundamental Physics, Zewail City of Science and Technology, \\
6 October City, Giza 12588, Egypt}
\author{Stefano Moretti} 
\email{S.Moretti@soton.ac.uk}
\affiliation{School of Physics and Astronomy, University of Southampton, \\
Southampton, SO17 1BJ, United Kingdom}



\begin{abstract}
In the context of the $B-L$ Supersymmetric Standard Model 
(BLSSM), we investigate the consistency of a light Higgs 
boson, with mass around $90-95$ GeV, with the results of a 
search performed by the CMS collaboration in the di-photon 
channel at the integrated luminosity of  35.9 fb$^{-1}$ and 
$\sqrt s$ = 13 TeV.
\end{abstract}



\maketitle


\section{Introduction}
\label{sec:intro}

The discovery of a Higgs boson compatible with the one predicted 
by the Standard Model (SM), $h$, with a mass of 125 GeV, at the 
Large Hadron Collider (LHC) in July 2012, has been considered 
as the beginning of a new era in particle physics. In fact, 
such a Higgs boson is the first fundamental (i.e., point-like) 
scalar particle (i.e., with spin 0 and CP-even) to be found in 
Nature and the last hitherto undiscovered object needed to 
complete the experimental verification of the SM. This 
detection confirmed the Higgs mechanism of Electro-Weak 
Symmetry Breaking (EWSB)  generating masses for fundamental 
particles. It also boosted the expectation of discovering New 
Physics (NP) Beyond the SM (BSM), as we also know that for, 
the aforementioned mass value, the SM is theoretically 
inconsistent. Many of the SM flaws (e.g., the hierarchy 
problem, the absence of coupling unification, etc.) can 
however be remedied by Supersymmetry (SUSY), although the  
latter has itself drawbacks (e.g., the $\mu$ problem, the poor 
consistency of a unified version of it with both collider and 
Dark Matter (DM) data, etc.) if formulated in its minimal 
version, the so-called Minimal Supersymmetric Standard Model 
(MSSM). However, non-minimal realisations of SUSY, e.g., with 
an enlarged gauge and/or Higgs sector, are both theoretically 
plausible and better compatible with experimental data
\cite{Book}. 

The statistically most significant channel leading to the 2012 
signal emerged in the $gg\to h \to \gamma\gamma$ production 
and decay mode, primarily thanks to the high experimental 
resolution that can be achieved (in the invariant mass of the 
two photons, $M_{\gamma\gamma}$) via the di-photon final state. 
Hence, it is not surprising that this channel is being routinely
used by ATLAS and CMS in their search for additional (neutral) 
Higgs bosons, an endeavour that has indeed started immediately 
after the aforementioned discovery, since most BSM scenarios 
(Supersymmetric and not) predict the existence of extra neutral 
Higgs states. The possibility of the existence of the latter, 
lighter or heavier than the SM state, thus is an open and 
challenging phenomenological problem.

The CMS collaboration has recently found potential signals 
for another neutral Higgs boson, $h^\prime$, with a mass of 90 
to 95 GeV, precisely in the discussed gluon-fusion initiated 
channel leading to the di-photon final state, i.e., 
$gg\to h'\to \gamma\gamma$. The corresponding data were 
collected at Center-of-Mass (CM) energies of $\sqrt{s} = 8$ 
and $13$ TeV and integrated luminosities of 19.7 and 35.9 
fb$^{-1}$, respectively \cite{Sirunyan:2018aui}. Based on 
these data, the CMS collaboration observed a resonant
structure at 90--95 GeV in the $M_{\gamma\gamma}$ spectrum with 
a local (global) significance of $2.8~(1.3)$ standard 
deviations, respectively. Despite the fact that this hint for 
a new resonance is still very preliminary, it has gained some 
attention in the particle physics community and several BSM 
explanations for it have been proposed \cite{BSM95}
If these observations are confirmed by  future data, it 
will be a significant direct evidence of NP.

In the present paper we show that a specific extension of the 
MSSM, the the so-called $B-L$ Supersymmetric SM (BLSSM) 
\cite{BLSSM}
which has a rich Higgs sector, consisting of two Higgs doublets 
and two Higgs singlets, can accommodate the observed anomaly. 
In particular, we emphasize that one of the CP-even Higgs bosons 
of this BSM construct can act as the potential $h^\prime$ state 
behind the aforementioned excess in $M_{\gamma\gamma}$, with 
the model still providing a SM-like Higgs state with 125 GeV, 
thus compatible with current LHC measurements. The BLSSM is an 
extension of the MSSM obtained by adopting an additional 
$U(1)_{B-L}$ gauge group, i.e., the full gauge structure is 
$SU(3)_C \times SU(2)_L \times U(1)_Y \times U(1)_{B-L}$. 
This model contains three SM singlet chiral superfields 
${\widehat N}_{1,2,3}$ (yielding right-handed neutrinos), two 
SM singlet chiral Higgs superfields ${\widehat \chi}_{1,2}$ 
(providing three additional physical Higgs states) and the 
${\widehat Z}^\prime$ vector superfield associated with the 
$U(1)_{B-L}$ gauge boson (embedding a physical $Z^\prime$ 
state), in addition to the MSSM superfields. Interestingly, 
it was shown that the scale of $B-L$ symmetry breaking is 
related to the soft SUSY-breaking scale \cite{Khalil:2007dr}, 
so that it is not unreasonable to find that this model can 
predict right-handed neutrinos, $Z'$ and Higgs states at or 
even  below the TeV scale. 

The mixing between the SM-like Higgs state $h$ and the 
BLSSM-specific Higgs state $h^\prime$ is proportional to the 
gauge coupling of the gauge kinetic mixing $\widetilde{g}$ 
between the $Z$ and $Z'$, which is (in a non-universal 
description) a free parameter and can be of order of 0.5. 
In this case, a large Higgs mixing is generated, which yields 
significant couplings between the $h^\prime$ and SM fermions 
and gauge bosons. Therefore, the production and decay rates 
of the $h^\prime$ state are not generally suppressed, 
including in the $gg\to h'\to\gamma\gamma$ channel, which 
proceeds mainly via top quark and $W^\pm$ gauge boson loops 
at production and decay level, respectively. Hence, the BLSSM 
can account for the observed 90--95 GeV potential signal.

The paper is organised as follows. In Sec.~\ref{sec:blssm}, 
we describe the Higgs sector of the BLSSM and emphasize that 
the mass of the lightest CP-even Higgs boson can naturally be 
around $90-95$ GeV with also a SM-like Higgs state having a 
mass of $125$ GeV. In Sec.~\ref{sec:results}, we investigate 
the would be BLSSM signal in the 
$gg\to h^\prime \to \gamma \gamma$ channel and show that 
it can explain the excesses presently observed by the CMS 
collaboration as well as offer a chance for $h'$ discovery 
already with the full Run 2 data set. Our conclusions are 
presented in Sec.~\ref{sec:conclude}.


\section{The BLSSM Higgs sector}
\label{sec:blssm}

The BLSSM superpotential is given by 
\begin{eqnarray}\label{eq:superp}
W_{\rm BLSSM} &=& y_u \widehat{Q} \widehat{H}_2 \widehat{U}^c 
+ y_d \widehat{Q} \widehat{H}_1 \widehat{D}^c 
+ y_e \widehat{L} \widehat{H}_1 \widehat{E}^c \nonumber \\ 
&+& \mu \widehat{H}_1 \widehat{H}_2 
+ y_\nu \widehat{L} \widehat{H}_2 \widehat N^c
+ y_N \widehat N^c \widehat{\chi}_1 \widehat N^c \nonumber \\ 
&+& \mu^\prime \widehat\chi_1 \widehat\chi_2\,,
\end{eqnarray}
where the first four terms are the usual MSSM ones, the next 
two terms represent the Yukawa interactions of the known 
neutrinos and between the additional right-handed ones $N_i$ 
($i=1,2,3$) and the singlet Higgs field $\chi_1$, respectively. 
The last term represents the bilinear mixing between $\chi_1$ 
and $\chi_2$. $y_u$, $y_d$, $y_e$ and $y_\nu$ are the quark, 
lepton and neutrino Yukawa coupling constants, respectively. 
Furthermore, $y_N$ is the the Yukawa coupling constant 
between $N_i$ and $\chi_1$,  $Q$ and $L$ are the left-handed 
quark and lepton doublet superfields while $U$, $D$ and $E$ 
are the right-handed up-type, down-type and electron-type 
singlet  ones, respectively. The charge conjugation is denoted 
by the superscript $c$. Then, $H_1$ and $H_2$ are the 
$SU(2)_L$ Higgs doublet superfields with opposite hypercharge 
$Y = \pm 1$.

One obtains the masses of the physical neutral BLSSM Higgs 
states in terms of the Higgs fields,
\begin{eqnarray}\label{eq:higgsfield}
H_{1,2}^0 = \frac{1}{\sqrt 2}(v_{1,2}+\sigma_{1,2}+i\phi_{1,2})\,,
\nonumber \\
\chi_{1,2}^0 = \frac{1}{\sqrt 2}(v_{1,2}^{\prime}
+\sigma_{1,2}^{\prime}+i\phi_{1,2}^{\prime})\,,
\end{eqnarray}
where the real and imaginary parts correspond to the CP-even (or
scalar) and the CP-odd (or pseudoscalar) Higgs states. $v_{1,2}$ 
and $v_{1,2}^{\prime}$ are the Vacuum Expectation Values (VEVs) 
of the Higgs fields $H_{1,2}$ and $\chi_{1,2}$, respectively. 
The CP-odd neutral Higgs mass-squared matrix at the tree-level 
in the basis $(\phi_1,\phi_2,\phi^{\prime}_1,\phi^{\prime}_2)$
is given by
\begin{equation}\label{eq:oddmatrix}
{\cal A}^2 =
\left(\begin{array}{cccc} 
B_\mu {\rm tan}\beta & B_\mu & 0 & 0 
\\ & \\ 
B_\mu & B_\mu {\rm cot}\beta & 0 & 0 
\\ & \\
0 & 0 & B_{\mu^\prime} {\rm tan}\beta^\prime & B_{\mu^\prime} 
\\ & \\
0 & 0 & B_{\mu^\prime} & B_{\mu^\prime} {\rm cot}\beta^\prime
\end{array} \right)\,,
\end{equation}
with  
\begin{eqnarray}\label{eq:Bmu}
B_\mu &=& - \frac{1}{8}
\biggl\{
-2{\widetilde g}g_{BL}v^{\prime 2}{\rm cos}2\beta^\prime
+4M_{H_1}^2-4M_{H_2}^2
\nonumber \\
&+& (g_1^2+{\widetilde g}^2+g_2^2)v^2{\rm cos 2\beta}
\biggr\}
{\rm tan 2\beta}\,, \nonumber \\
B_{\mu^\prime} &=& - \frac{1}{4}
\biggl(
-2g_{BL}^2v^{\prime 2}{\rm cos}2\beta^\prime
+2M_{\chi_1}^2-2M_{\chi_2}^2
\nonumber \\
&+& {\widetilde g}g_{BL}v^2{\rm cos 2\beta}
\biggr)
{\rm tan 2\beta^\prime}\,,
\end{eqnarray}
where tan$\beta=\frac{v_2}{v_1}$ and 
tan$\beta^\prime=\frac{v^\prime_2}{v^\prime_1}$.
$g_{BL}$ is the gauge coupling constant of $U(1)_{B-L}$ and 
$\widetilde g$ is the gauge coupling constant of the mixing 
between $U(1)_Y$ and $U(1)_{B-L}$. $g_1$ and $g_2$ are the 
$U(1)_Y$ and $SU(2)_L$ gauge coupling constants, respectively.

The CP-even neutral Higgs mass-squared matrix at the tree-level 
in the basis 
$(\sigma_1,\sigma_2,\sigma^{\prime}_1,\sigma^{\prime}_2)$ 
is given by
\begin{equation}\label{eq:evenmatrix}
{\cal M}^2 =
\left( \begin{array}{cc} {\cal M}_{hH}^2 & {\cal M}_{h h^\prime}^2 
\\ & \\
\left({\cal M}_{h h^\prime}^2\right)^T & {\cal M}_{h^\prime H^\prime}^2 
\end{array} \right)\,,
\end{equation}
where ${\cal M}_{hH}$ is the MSSM CP-even mass matrix 
which results into a SM-like Higgs boson $h$ with a mass 
$m_h \sim 125$ GeV and a heavy Higgs boson $H$ with a mass 
$m_H 
\sim {\cal O}(1~{\rm TeV})$. The BLSSM mass 
matrix ${\cal M}_{h^\prime H^\prime}$ reads 
\small{
\begin{equation}\label{eq:blssmmatrix}
{\cal M}_{h^\prime H^\prime}^2 =
\left(\begin{array}{cc} 
m_{A^\prime}^2 c_{\beta^\prime}^2 
+ g_{BL}^2 v_1^{\prime 2}
& 
-\frac{1}{2}m_{A^\prime}^2 s_{2\beta^\prime}
- g_{BL}^2 v_1^{\prime} v_2^{\prime} 
\\ & \\
-\frac{1}{2}m_{A^\prime}^2 s_{2\beta^\prime}
- g_{BL}^2 v_1^{\prime} v_2^{\prime} 
& 
m_{A^\prime}^2 s_{\beta^\prime}^2 
+ g_{BL}^2 v_2^{\prime 2}
\end{array}\right)
\end{equation}}
\noindent
with $c_x={\rm cos}x$ and $s_x={\rm sin}x$. 

The CP-even physical Higgs mass states can be obtained by 
diagonalising the Higgs mass-squared matrix given by 
Eq.~(\ref{eq:evenmatrix}) with a unitary matrix ${\cal R}$ as follows:
\begin{equation}\label{eq:diagonalise}
{\cal R} {\cal M}^2 {\cal R}^\dagger = {\rm diag}
\{m_h^2,m_{h^\prime}^2,m_H^2,m_{H^\prime}^2\}\,.
\end{equation}

\begin{figure}
\includegraphics[scale=0.7]{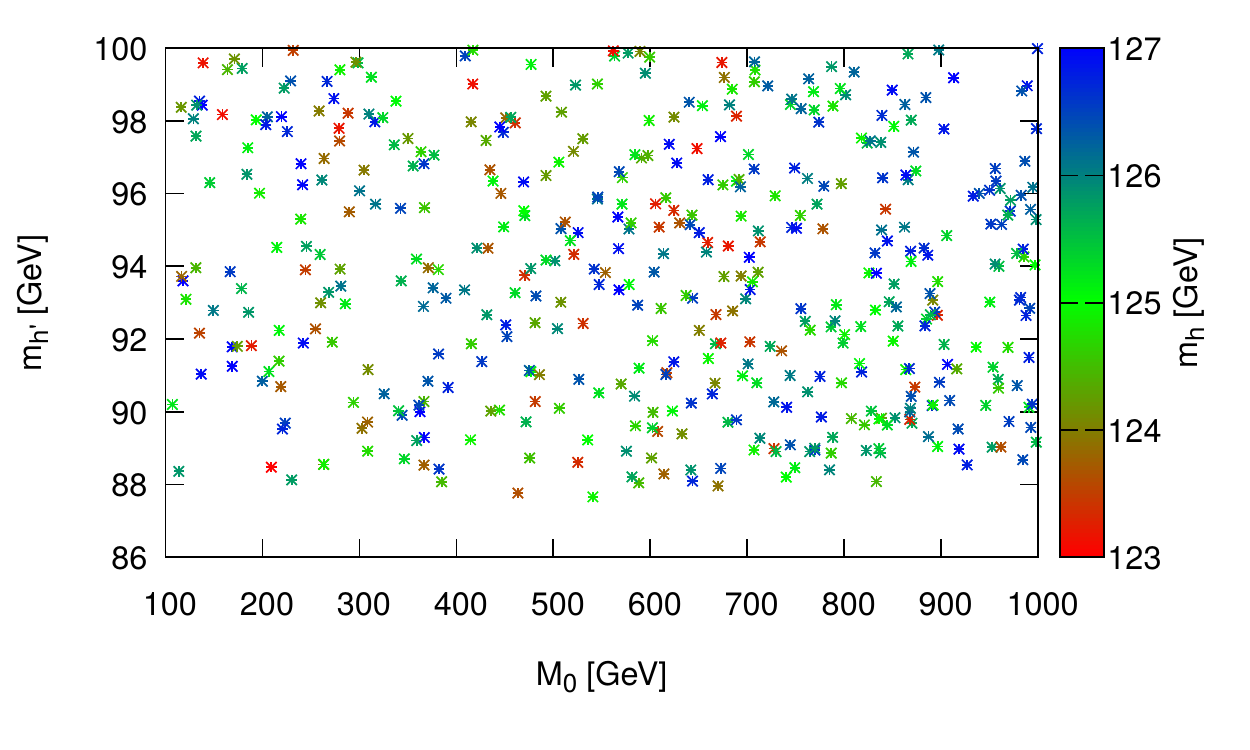}
\caption{\label{fig:m0-mh1}Scan of $m_{h^\prime}$ vs  $M_0$ with $m_h$ as a colour map.}
\end{figure}
\begin{figure}
\includegraphics[scale=0.7]{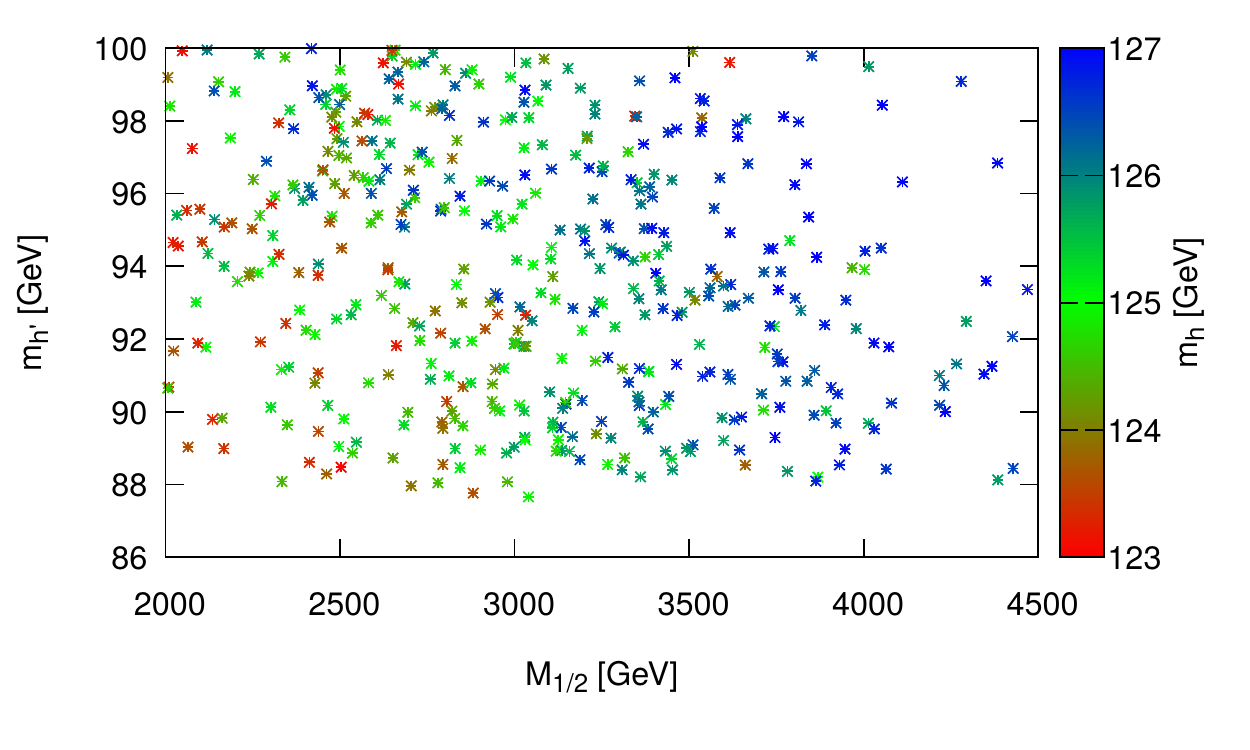}
\caption{\label{fig:m12-mh1}Scan of $m_{h^\prime}$ vs $M_{\frac{1}{2}}$ with $m_h$ as a colour map.}
\end{figure}

In order to find solutions consistent with the CMS observation 
of a scalar of mass around 90--95 GeV~\cite{Sirunyan:2018aui}, 
we perform a parameter space scan in the BLSSM with an in-house 
scanning tool which calls the public spectrum generator 
{\tt SPheno-v4.0.4} \cite{SPheno}
to generate the particle spectrum for each randomly scanned 
parameter space point. {\tt SPheno} requires the model files 
to generate the output spectrum in the context of a particular 
model (in our case it is the BLSSM) for a given point.
These model files are generated with the public package 
{\tt SARAH-v4.14.3}~\cite{Staub:2015kfa}. We perform the scan 
at the GUT scale by varying four input parameters, namely, the 
universal Soft SUSY-Breaking (SSB) scalar mass term 
$M_0~(=M_{Q_{1,2,3}}=M_{U_{1,2,3}}=M_{D_{1,2,3}}
=M_{L_{1,2,3}}=M_{E_{1,2,3}})$, the universal SSB gaugino 
mass term $M_{\frac{1}{2}}~(=2M_1=M_2=\frac{1}{3}M_3)$, 
tan$\beta$ and the universal Higgs to sfermion trilinear coupling 
$A_0~(=A_{\widetilde t}=A_{\widetilde b}=A_{\widetilde \tau})$,
while keeping all other model parameters fixed, e.g., 
$m_{Z^{\prime}}=2500$ GeV, tan$\beta^{\prime}=1.15$ and 
$\mu=\mu^{\prime}=B_{\mu}=B_{\mu^{\prime}}=0$. The ranges for 
the variable input parameters are given in Tab. I.

\begin{table}\label{tab:param}
\vskip 0.5cm 
\centering\begin{tabular}{|c|c|} 
\hline
Parameter & Range \\ 
\hline
$M_0$ & 100 -- 1000 GeV\\ 
\hline
$M_{\frac{1}{2}}$ & 1000 -- 4500 GeV \\
\hline
tan$\beta$ & 1 -- 60 \\
\hline
$A_0$ & 1000 -- 4000 GeV \\
\hline
\end{tabular}
\caption{Ranges for the four variable input parameters.}
\end{table}

The randomly scanned points are required to produce the lightest 
neutral Higgs boson mass in the range 90 GeV 
$\ge m_{h^{\prime}}\ge$ 95 GeV (approximately). As far as the 
experimental constraints are concerned, these points should also 
result in a SM-like Higgs boson with a mass $m_h$ 
which allows $\pm 2$ GeV uncertainty in its theoretical model 
prediction, 
about the experimental measurement of 
$m_h=125.09 \pm 0.32$ GeV~\cite{Aad:2015zhl}. Moreover, 
the points are passed through {\tt HiggsBounds-v4.3.1}
\cite{HBounds} and
{\tt HiggsSignals-v1.4.0} \cite{HSignals}
to be consistent with the Higgs boson experimental measurements 
performed by the LEP, TeVatron and the LHC. {\tt SPheno} also 
calculates flavour observables, so that the scanned points 
also need to satisfy the experimental constraints on the 
Branching Ratios (BRs) of the most stringent $B$-meson decay 
channels within a 2$\sigma$ error, which are given by
${\rm BR}(B\to X_s \gamma) =
(3.32\pm0.15) \times 10^{-4}$,
${\rm BR}(B_s \to \mu^+ \mu^-) =
(3.1\pm 0.6) \times 10^{-9}$ and 
${\rm BR}(B_u\to \tau^\pm \nu_\tau) =
(1.06\pm0.19) \times 10^{-4}$ \cite{Amhis:2019ckw}.

In Fig.~\ref{fig:m0-mh1} we present our randomly scanned 
points on the $M_0-m_{h^\prime}$ plane where the colour map 
represents the values of $m_h$ for those points. It shows a 
good possibility of having BLSSM solutions with a light scalar 
state of mass of $90-95$ GeV and a SM-like scalar with a mass 
near 125 GeV, at the same time. Similarly, Fig. 
\ref{fig:m12-mh1} depicts the scanned points on the 
$M_{\frac{1}{2}}-m_{h^\prime}$ plane while the colour map 
shows the values of $m_h$. Note that the points with 
$M_{\frac{1}{2}} < 2000$ GeV are excluded by {\tt HiggsBounds}.

In the next section, we present our Monte Carlo (MC) analysis
in the light of the CMS observation of a light scalar in terms
of a few Benchmark Points (BPs) selected from our random scan. 
The details of these BPs are listed in Tab.~\ref{tab:bps}. Note 
that the tabulated cross sections (given at 13 TeV) are 
calculated with the public package {\tt MadGraph5-v1.5.1}
\cite{Alwall:2011uj}, which is also used for our (irreducible) 
background, i.e., $q\bar q, gg\to \gamma\gamma$\footnote[1]{As 
the majority of the excess in the CMS analysis comes from the 
higher energy data, henceforth, we neglect benchmarking against 
the 8 TeV ones.}. The ensuing Leading Order (LO) results are 
supplemented by inclusive $k$-factors for both signals and 
background, as follows. We consider the Next-to-LO (NLO) 
$k$-factor which is defined as 
$k_{\rm NLO}=\frac{\sigma_{\rm NLO}}{\sigma_{\rm LO}}$. For 
the signal, in order to estimate $k_{\rm NLO}$, we calculate 
$\sigma(gg \to h,h^{\prime})$ both at both LO and NLO using 
the public tool {\tt SusHi-v1.7.0}~\cite{Harlander:2012pb},  
since here the largest higher order corrections are only 
associated with the production process. Here, the value of 
$k_{\rm NLO}$ is essentially $2.4$ in the entire mass range 
$90-125$ GeV. For the background, we assume a constant 
$k_{\rm NLO}=1.3$ in our analysis, following Ref. 
\cite{Catani:2018krb}.

\begin{table}\label{tab:bps}
\vskip 0.5cm
\centering\begin{tabular}{|c|c|c|c|c|}
\hline
BP & $m_{h^\prime}$ & $m_h$ & 
$\sigma(pp \to h^\prime \to \gamma \gamma)$ &  
$\sigma(pp \to h \to \gamma \gamma)$ \\
\hline
1 & 95.3 & 125.9 & 13.1 & 43.5 \\
\hline
2 & 94.2 & 125.3 & 8.6 & 49.3 \\
\hline
3 & 89.7 & 125.7 & 9.7 & 49.3 \\
\hline
4 & 90.0 & 127.2 & 8.7 & 47.6 \\
\hline
\end{tabular}
\caption{The masses (in GeV) of the two lightest neutral Higgs 
bosons and the cross sections (in fb) at 13 TeV for the processes
$pp \to h^\prime \to \gamma \gamma$ and 
$pp \to h \to \gamma \gamma$ for the BPs presented in 
Sec.~\ref{sec:results}.}
\end{table}


\section{Numerical analysis and results}
\label{sec:results}

\begin{figure}
\includegraphics[scale=0.40]{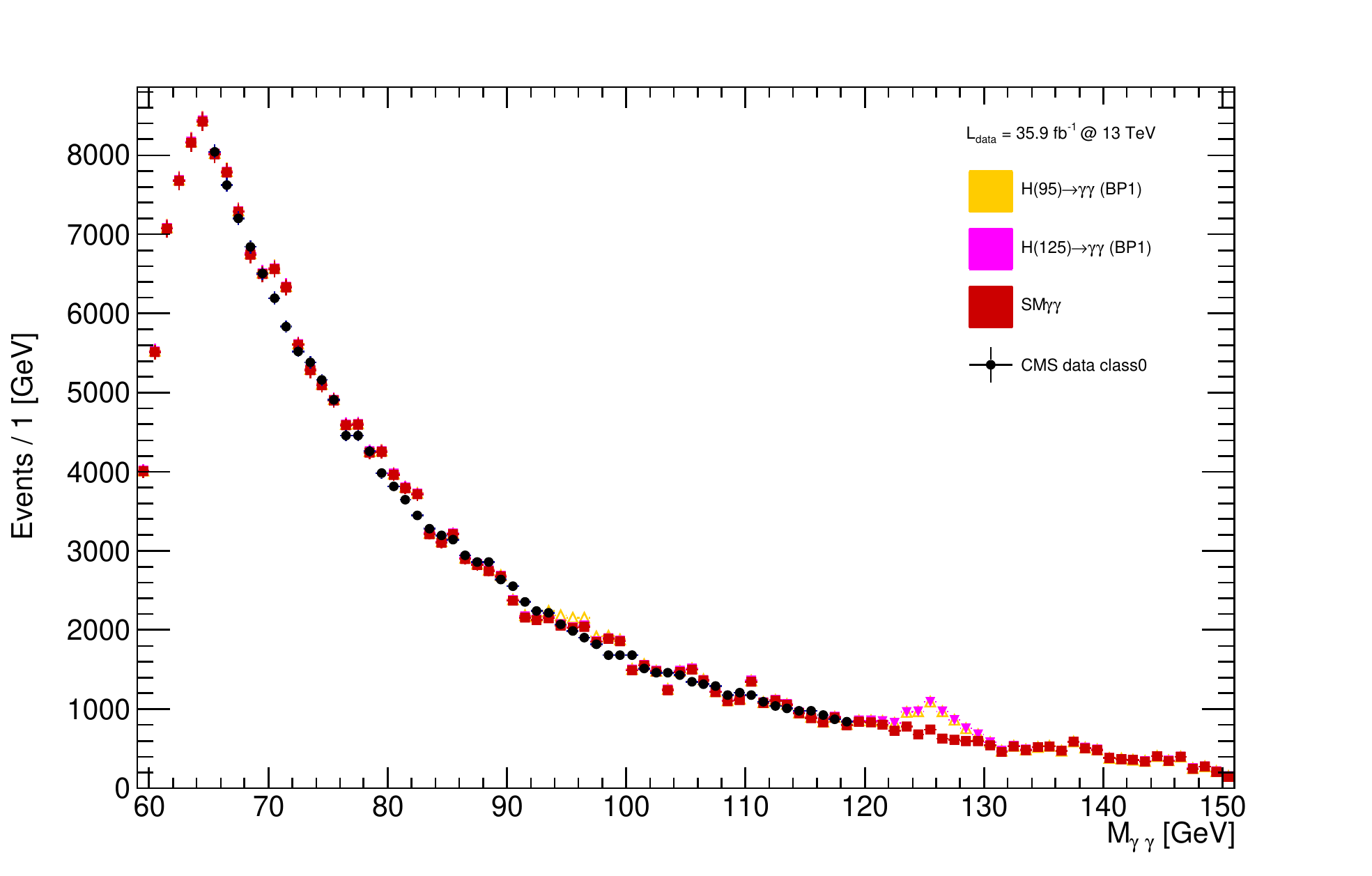}
\caption{\label{fig:class0vsBP1_13TeV}BP1 versus CMS data at 
13 TeV~\cite{Sirunyan:2018aui}. Yellow points represent 
$h^\prime \to \gamma \gamma$, pink points represent 
$h \to \gamma \gamma$ while red points show the SM background.}
\end{figure}
\begin{figure}
\vskip 5mm
\includegraphics[scale=0.40]{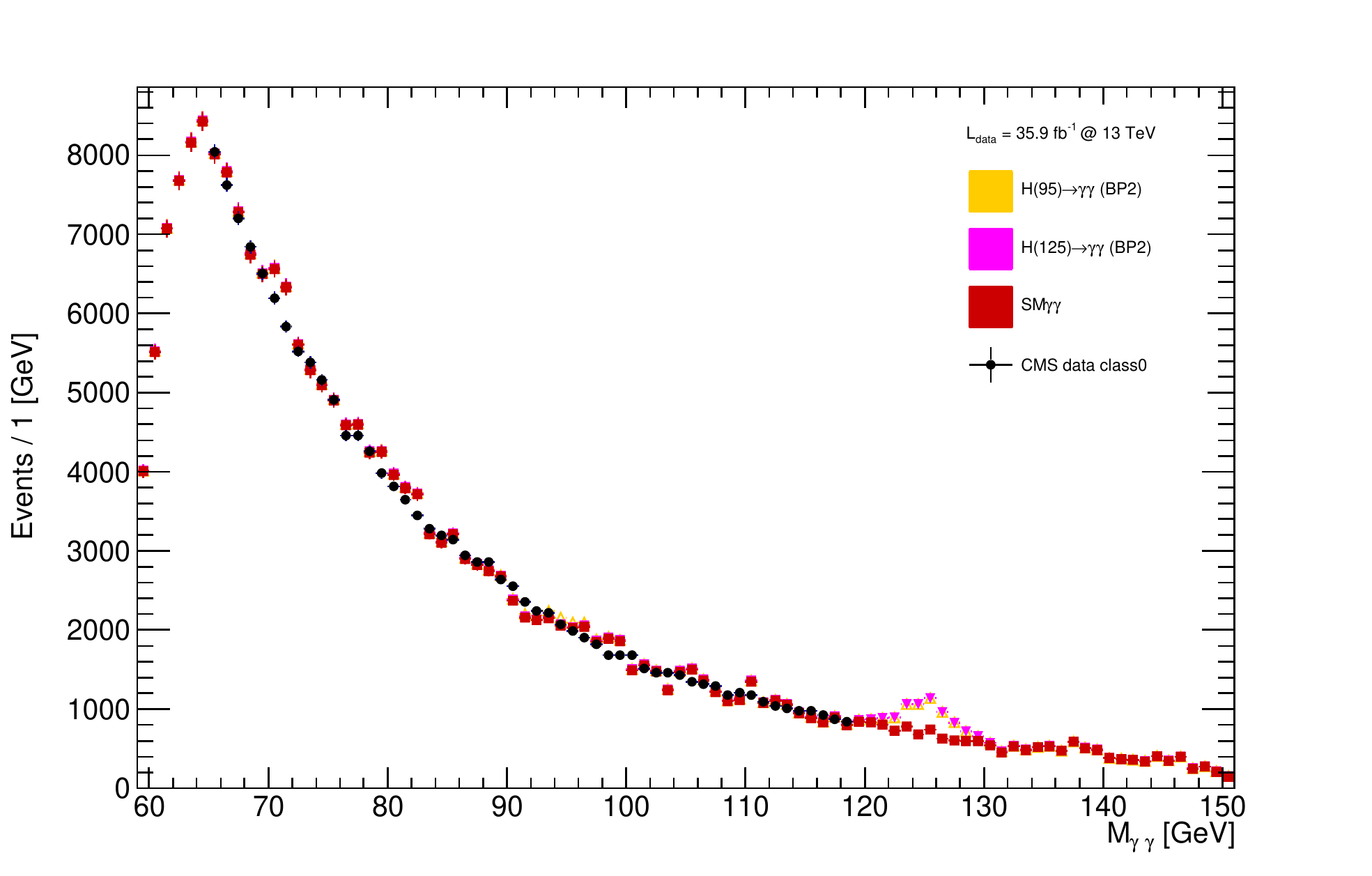}
\caption{\label{fig:class0vsBP3_13TeV}BP2 versus CMS data at 
13 TeV~\cite{Sirunyan:2018aui}. Yellow points represent 
$h^\prime \to \gamma \gamma$, pink points represent 
$ h \to \gamma \gamma$ while red points show the SM background.}
\end{figure}
\begin{figure}
\includegraphics[scale=0.40]{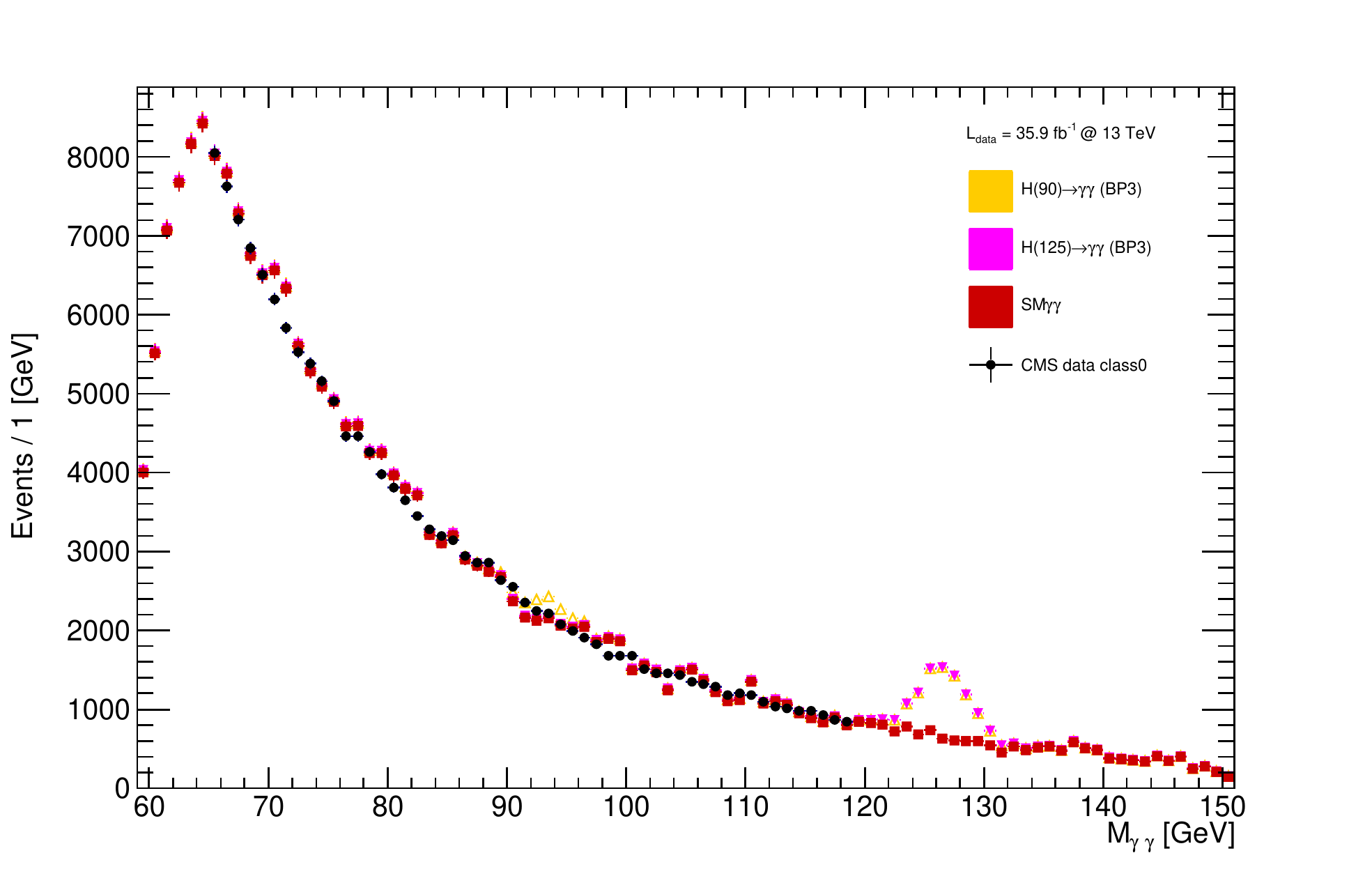}
\caption{\label{fig:class0vsh90BP1_13TeV}BP3 versus CMS data 
at 13 TeV~\cite{Sirunyan:2018aui}. Yellow points represent 
$h^\prime \to \gamma \gamma$, pink points represent 
$h \to \gamma \gamma$ while red points show the SM background.}
\end{figure}
\begin{figure}
\includegraphics[scale=0.40]{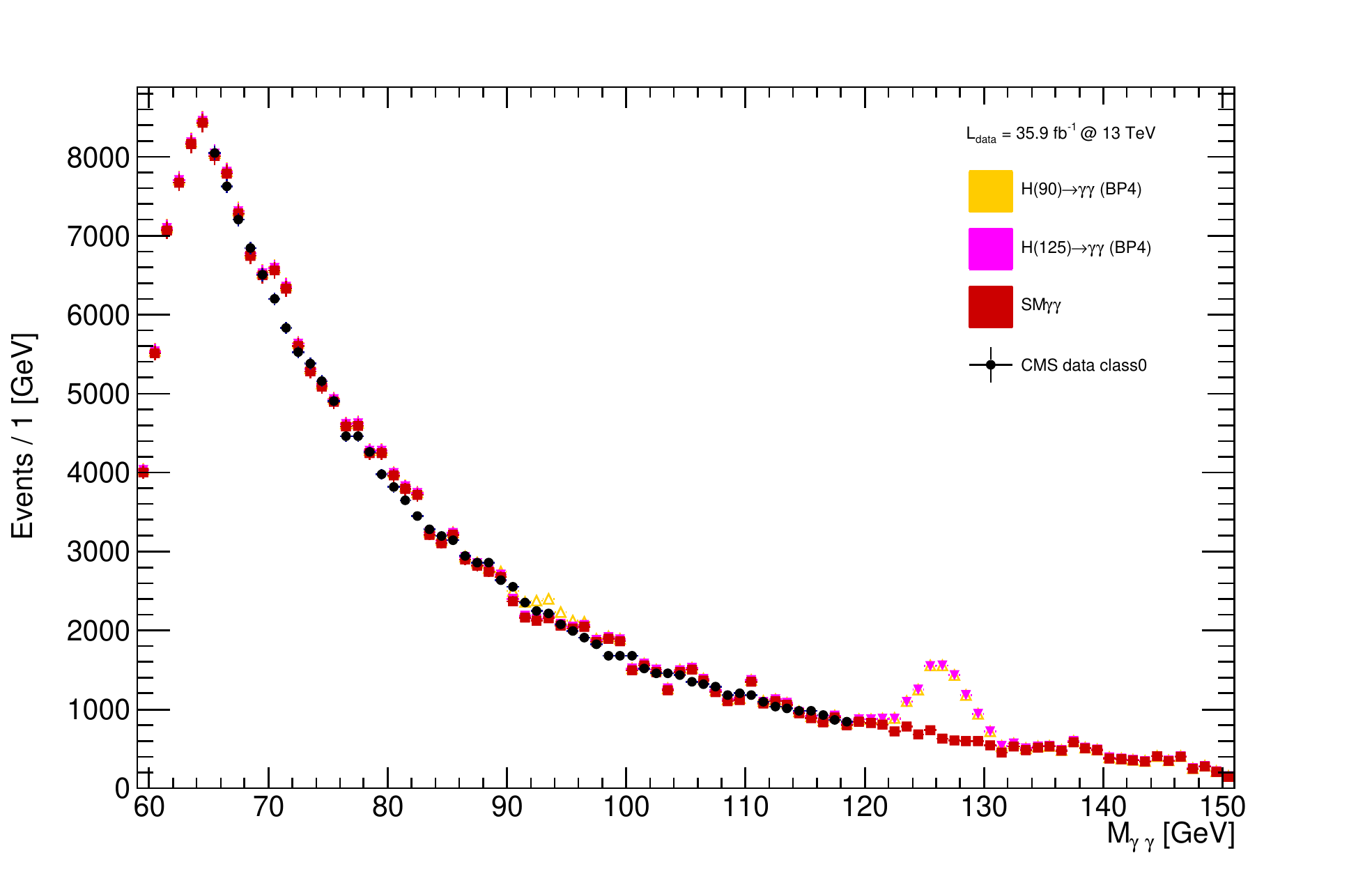}
\caption{\label{fig:class0vsh90BP2_13TeV}BP4 versus CMS data 
at 13 TeV~\cite{Sirunyan:2018aui}. Yellow points represent 
$h^\prime \to \gamma \gamma$, pink points represent 
$h \to \gamma \gamma$ while red points show the SM background.}
\end{figure}
\begin{table}\label{tab:eventYield}
\vskip 0.5cm
\centering\begin{tabular}{|c|c|c|c|c|}
\hline
Range of $M_{\gamma\gamma}$ [GeV] & [65-119] & [85-100] & [92-98] & [89-98] \\
\hline
CMS data & 170019  & 38159 & 14608 & 22654 \\
\hline
SM & 171337 & 37986 & 14414 & 22202 \\
\hline
$h^\prime$ (BP1) & 726  & 605 & 536 & -- \\
$h$ (BP1) & 549  & 167 & 72 & -- \\
$h^\prime$+$h$+SM (BP1) & 172612 & 38758 & 15022 & -- \\
\hline
$h^\prime$ (BP2) & 472  & 396 & 356 & -- \\
$h$ (BP2) & 633 & 192 & 82 & -- \\
$h^\prime$+$h$+SM (BP2) & 172442  & 38574 & 14852 & -- \\
\hline
$h^\prime$ (BP3) & 1421  & 1196 & -- & 1150 \\
$h$ (BP3) & 1380  & 419 &--  & 261 \\
$h^\prime$+$h$+SM (BP3) & 174138  & 39601 &--  &  23613 \\
\hline
$h^\prime$ (BP4) & 1305  & 1098  & -- &1057  \\
$h$ (BP4) & 1422  & 431 & -- & 266 \\
$h^\prime$+$h$+SM (BP4) & 174064  & 39515 & -- & 23525 \\
\hline
\end{tabular}
\vskip 5mm
\caption{Number of events in the CMS data of 
\cite{Sirunyan:2018aui} and our MC samples in different 
$M_{\gamma\gamma}$ ranges. Note that the empty cells in the 
table were not used in the significance calculation.}
\end{table}
\begin{figure}
\vspace*{-3truemm}
\includegraphics[scale=0.40]{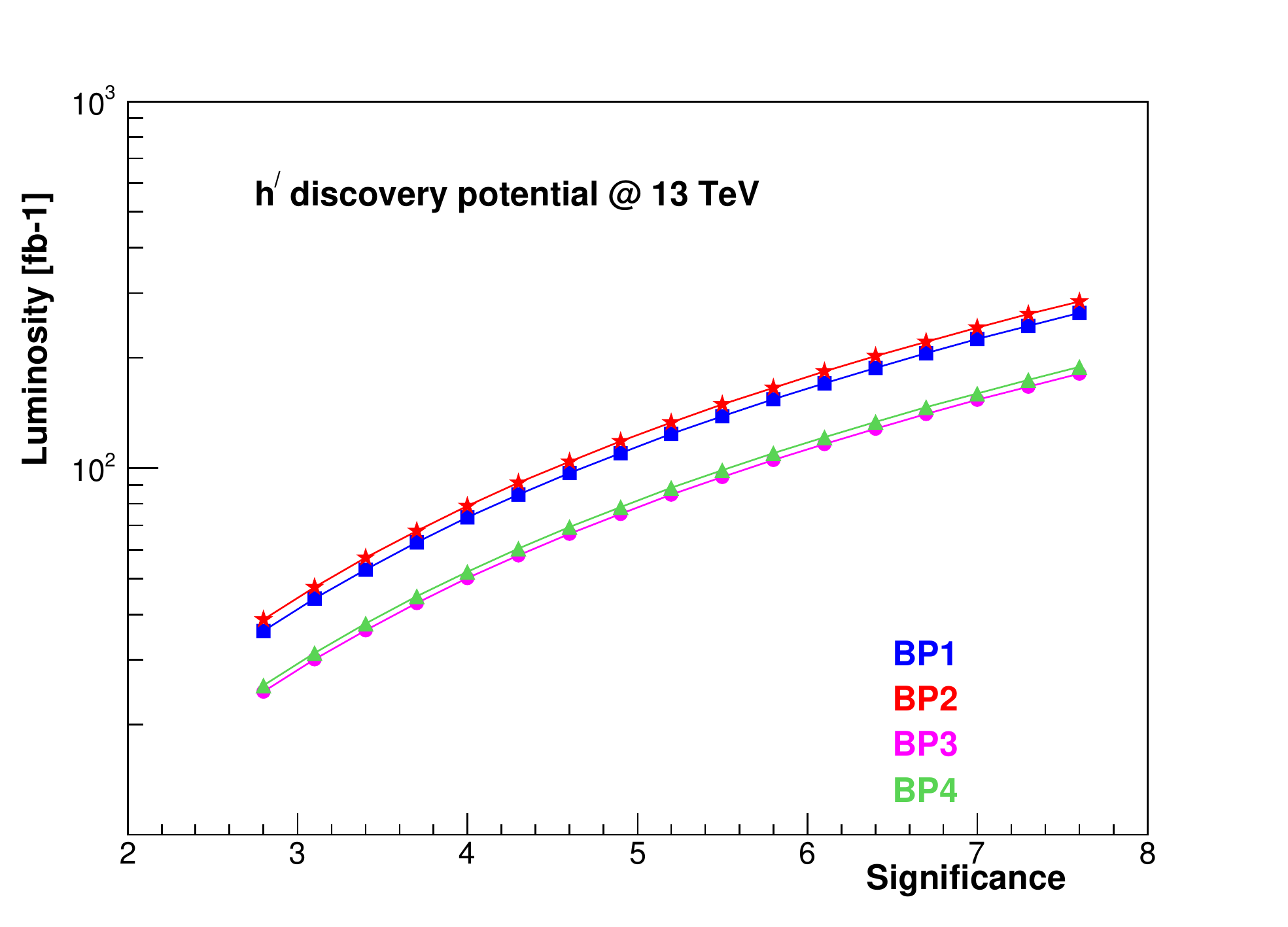}
\caption{\label{fig:h1_discoveryPotential_13TeV}Integrated 
luminosity needed for $h^\prime$ discovery in the di-photon 
channel as a function of significance for BP1--4.}
\end{figure}

In our analysis, we selected all events that contain a 
di-photon pair in the detector fiducial range 
$|\eta^\gamma | \leqslant 2.5 $ and out of the crack region 
between the barrel and end-cap parts of the CMS 
Electro-Magnetic (EM) calorimeters. Each photon in the pair 
has to satisfy a requirement on the ratio of its 
$p_T^{\gamma_i}$ ($i=1,2$, with 1(2) being the most(least) 
energetic one) value to the invariant mass of the di-photon 
system. These requirements are 
$p^{\gamma_1}_T / M_{\gamma \gamma} > 30.6/65.0 = 0.47$ and 
$p^{\gamma_2}_T / M_{\gamma \gamma} > 18.2/65.0 = 0.28$. Our 
results are therefore directly comparable to the CMS class0 
data of \cite{Sirunyan:2018aui}, which apply the same 
requirements on the di-photon system. We have digitised such 
data (see black cross symbols thereafter). 
Fig.~\ref{fig:class0vsBP1_13TeV}--\ref{fig:class0vsBP3_13TeV}(\ref{fig:class0vsh90BP1_13TeV}--\ref{fig:class0vsh90BP2_13TeV}) 
shows the $M_{\gamma \gamma}$ distribution for the aforementioned CMS data (at 13 
TeV)  alongside the MC ones for our BP1--2(BP3--4), where yellow markers refer to the $h^\prime$ signal, 
pink markers refer to the $h$ signal while red markers refer 
to the SM background (the former two being stacked onto the latter). In Figs.~\ref{fig:class0vsBP1_13TeV}--\ref{fig:class0vsBP3_13TeV}(\ref{fig:class0vsh90BP1_13TeV}--\ref{fig:class0vsh90BP2_13TeV}), 
we see moderate peaks stemming from the background for the 
$h^\prime$ signals around 95(90) GeV and clear peaks for the 
$h$ ones around 125 GeV. To convince oneself of the statistical 
relevance of both Higgs boson peaks, we present in 
Tab. \ref{tab:eventYield} a comparison between the number of 
events from each signal and background. We used the number of 
events in the di-photon mass range $92-98$ GeV($89-98$ GeV) 
to calculate the significance for BP1--2(BP3--4). This  
 has been calculated using the formula 
$S/\sqrt{B}$, where $S$ is the number of $h^\prime$ events 
and $B$ is that of background ones.  

Finally, Fig.~\ref{fig:h1_discoveryPotential_13TeV} shows the 
integrated luminosity needed to discover the $h^\prime$ state 
of the BLSSM  in di-photon events using CMS data at 13 TeV for 
our four BPs. It is clear that, for all of the latter, 
discovery is within reach of Run 2, as luminosity values of 
114(123)[79]\{82\} fb$^{-1}$ are needed to reach a $5\sigma$ 
excess in the $90-95$ GeV region for BP1(2)[3]\{4\}.


\section{Conclusions}
\label{sec:conclude}

Motivated by a $\sim 2.8\sigma$ excess recorded by the CMS experiment  in the di-photon channel at the integrated luminosity of  35.9 fb$^{-1}$ at  $\sqrt s$ = 13 TeV (in fact, with a moderate contribution from 8 TeV data too) around a mass of order $ 90-95$ GeV, we have analysed the discovery potential of a light neutral Higgs boson $h'$ available  in the context of the BLSSM at Run 2 of the LHC. 
We considered four BPs and showed that each of these can produce an enhancement of the di-photon cross section in the above mass region through the sub-process  $gg \to h' \to \gamma \gamma$ compatible with the CMS anomalous data while simultaneously producing the required amount of signal induced in the same channel by the SM-like state of the BLSSM, so as to comply with the di-photon data collected around 125 GeV.   We also estimated the required integrated luminosity needed for a $5\sigma$ discovery of such 
$h'$ state in the above channel, which turned out to be less than the total Run 2 data sample, so that we advocate new analyses using the latter.

\vspace*{-0.5truecm}
\section*{Acknowledgments}
\vspace*{-0.25truecm}
BD acknowledges the financial support provided by ICTP-EAIFR 
where part of this project was carried out. SM is financed in 
part through the NExT Institute and STFC Consolidated Grant 
No. ST/L000296/1.


\begin{thebibliography}{99}


\bibitem{Book} S. Khalil and S. Moretti, {\sl Supersymmetry Beyond Minimality: from Theory to Experiment}
(CRC Press, Taylor \& Francis Group, Boca Raton, FL, 2019).

\bibitem{Sirunyan:2018aui}
A.~M.~Sirunyan \textit{et al.} [CMS],
\href{http://dx.doi.org/10.1016/j.physletb.2019.03.064}
{Phys. Lett. B \textbf{793}, 320 (2019)}.

\bibitem{BSM95}
T.~Biek\"otter, M.~Chakraborti and S.~Heinemeyer,
\href{http://arxiv.org/abs/2003.05422}
{arXiv:2003.05422 [hep-ph]};
J.~A.~Aguilar-Saavedra and F.~R.~Joaquim,
\href{http://dx.doi.org/10.1140/epjc/s10052-020-7952-4}
{Eur. Phys. J. C \textbf{80}, no.5, 403 (2020)};
T.~Biek\"otter, M.~Chakraborti and S.~Heinemeyer,
\href{http://arxiv.org/abs/1910.06858}
{arXiv:1910.06858 [hep-ph]};
J.~Cao, X.~Jia, Y.~Yue, H.~Zhou and P.~Zhu,
\href{http://dx.doi.org/10.1103/PhysRevD.101.055008}
{Phys. Rev. D \textbf{101}, no.5, 055008 (2020)};
K.~Choi, S.~H.~Im, K.~S.~Jeong and C.~B.~Park,
\href{http://dx.doi.org/10.1140/epjc/s10052-019-7473-1}
{Eur. Phys. J. C \textbf{79}, no.11, 956 (2019)};
T.~Biek\"otter, M.~Chakraborti and S.~Heinemeyer,
\href{http://dx.doi.org/10.22323/1.347.0015}
{PoS \textbf{CORFU2018}, 015 (2019)};
T.~Biek\"otter, M.~Chakraborti and S.~Heinemeyer,
\href{http://dx.doi.org/10.1140/epjc/s10052-019-7561-2}
{Eur. Phys. J. C \textbf{80}, no.1, 2 (2020)};
S.~Heinemeyer and T.~Stefaniak,
\href{http://dx.doi.org/10.22323/1.339.0016}
{PoS \textbf{CHARGED2018}, 016 (2019)};
S.~Heinemeyer,
\href{http://dx.doi.org/10.1142/S0217751X18440062}
{Int. J. Mod. Phys. A \textbf{33}, no.31, 1844006 (2018)}.

\bibitem{BLSSM}
S.~Khalil and S.~Moretti,
\href{http://dx.doi.org/10.3389/fphy.2013.00010}
{Front. in Phys. \textbf{1}, 10 (2013)};
L.~Basso, A.~Belyaev, S.~Moretti, G.~M.~Pruna and C.~H.~Shepherd-Themistocleous,
\href{http://dx.doi.org/10.22323/1.084.0242}
{PoS \textbf{EPS-HEP2009}, 242 (2009)};
L.~Basso, A.~Belyaev, S.~Moretti and G.~M.~Pruna,
\href{http://dx.doi.org/10.1088/1742-6596/259/1/012062}
{J. Phys. Conf. Ser. \textbf{259}, 012062 (2010)};
L.~Basso, S.~Moretti and G.~M.~Pruna,
\href{http://dx.doi.org/10.1103/PhysRevD.83.055014}
{Phys. Rev. D \textbf{83}, 055014 (2011)};
W.~Emam and S.~Khalil,
\href{http://dx.doi.org/10.1140/epjc/s10052-007-0411-7}
{Eur. Phys. J. C \textbf{52}, 625 (2007)};
S.~Khalil,
\href{http://dx.doi.org/10.1088/0954-3899/35/5/055001}
{J. Phys. G \textbf{35}, 055001 (2008)}.

\bibitem{Khalil:2007dr}
S.~Khalil and A.~Masiero,
\href{http://dx.doi.org/10.1016/j.physletb.2008.06.063}
{Phys. Lett. B \textbf{665}, 374 (2008)}.

\bibitem{SPheno}
W.~Porod,
\href{http://dx.doi.org/10.1016/S0010-4655(03)00222-4}
{Comput. Phys. Commun. \textbf{153}, 275 (2003)};
W.~Porod and F.~Staub,
\href{http://dx.doi.org/10.1016/j.cpc.2012.05.021}
{Comput. Phys. Commun. \textbf{183}, 2458 (2012)}.

\bibitem{Staub:2015kfa}
F.~Staub,
\href{http://dx.doi.org/10.1155/2015/840780}
{Adv. High Energy Phys. \textbf{2015}, 840780 (2015)}.

\bibitem{Aad:2015zhl}
G.~Aad \textit{et al.} [ATLAS and CMS],
\href{http://dx.doi.org/10.1103/PhysRevLett.114.191803}
{Phys. Rev. Lett. \textbf{114}, 191803 (2015)}.

\bibitem{HBounds}
P.~Bechtle, O.~Brein, S.~Heinemeyer, G.~Weiglein and K.~E.~Williams,
\href{http://dx.doi.org/10.1016/j.cpc.2009.09.003}
{Comput. Phys. Commun. \textbf{181}, 138 (2010)};
P.~Bechtle, O.~Brein, S.~Heinemeyer, G.~Weiglein and K.~E.~Williams,
\href{http://dx.doi.org/10.1016/j.cpc.2011.07.015}
{Comput. Phys. Commun. \textbf{182}, 2605 (2011)};
P.~Bechtle, O.~Brein, S.~Heinemeyer, O.~St\r{a}l, T.~Stefaniak, G.~Weiglein and K.~E.~Williams,
\href{http://dx.doi.org/10.1140/epjc/s10052-013-2693-2}
{Eur. Phys. J. C \textbf{74}, no.3, 2693 (2014)}.

\bibitem{HSignals}
P.~Bechtle, S.~Heinemeyer, O.~St\r{a}l, T.~Stefaniak and G.~Weiglein,
\href{http://dx.doi.org/10.1140/epjc/s10052-013-2711-4}
{Eur. Phys. J. C \textbf{74}, no.2, 2711 (2014)};
O.~St\r{a}l and T.~Stefaniak,
\href{http://dx.doi.org/10.22323/1.180.0314}
{PoS \textbf{EPS-HEP2013}, 314 (2013)};
P.~Bechtle, S.~Heinemeyer, O.~St\r{a}l, T.~Stefaniak and G.~Weiglein,
\href{http://dx.doi.org/10.1007/JHEP11(2014)039}
{JHEP \textbf{11}, 039 (2014)}.

\bibitem{Amhis:2019ckw}
Y.~S.~Amhis \textit{et al.} [HFLAV],
\href{http://arxiv.org/abs/1909.12524}
{arXiv:1909.12524 [hep-ex]}.

\bibitem{Alwall:2011uj}
J.~Alwall, M.~Herquet, F.~Maltoni, O.~Mattelaer and T.~Stelzer,
\href{http://dx.doi.org/10.1007/JHEP06(2011)128}
{JHEP \textbf{06}, 128 (2011)}.

\bibitem{Harlander:2012pb}
R.~V.~Harlander, S.~Liebler and H.~Mantler,
\href{http://dx.doi.org/10.1016/j.cpc.2013.02.006}
{Comput. Phys. Commun. \textbf{184}, 1605 (2013)}.

\bibitem{Catani:2018krb}
S.~Catani, L.~Cieri, D.~de Florian, G.~Ferrera and M.~Grazzini,
\href{http://dx.doi.org/10.1007/JHEP04(2018)142}
{JHEP \textbf{04}, 142 (2018)}.

\end{thebibliography}
\end{document}